\documentclass[12pt,preprint]{aastex}

\shorttitle{The Binary White Dwarf LHS 3236}
\shortauthors{Harris et al.}

\begin{document}

\title{The Binary White Dwarf LHS 3236\altaffilmark{1}}

\author{Hugh C. Harris,\altaffilmark{2}
Conard C. Dahn,\altaffilmark{2}
Trent J. Dupuy,\altaffilmark{3,4,5}
Blaise Canzian,\altaffilmark{2,6}
Harry H. Guetter,\altaffilmark{2}
William I. Hartkopf,\altaffilmark{7}
Michael J. Ireland,\altaffilmark{8,9}
S.K. Leggett,\altaffilmark{10}
Stephen E. Levine, \altaffilmark{2,11}
Michael C. Liu,\altaffilmark{3,12}
Christian B. Luginbuhl, \altaffilmark{2}
Alice K. B. Monet, \altaffilmark{2,7}
Ronald C. Stone,\altaffilmark{2,13}
John P. Subasavage,\altaffilmark{2}
Trudy Tilleman,\altaffilmark{2}
Richard L. Walker\altaffilmark{2,13}
}

\altaffiltext{1}{Some of the data presented herein were obtained at
  the W.M. Keck Observatory, which is operated as a scientific
  partnership among the California Institute of Technology, the
  University of California, and the National Aeronautics and Space
  Administration. The Observatory was made possible by the generous
  financial support of the W.M. Keck Foundation.}
\altaffiltext{2}{US Naval Observatory,
       10391 West Naval Observatory Road,\\
       Flagstaff, AZ 86001-8521; hch@nofs.navy.mil}
\altaffiltext{3}{Institute for Astronomy, University of Hawaii, 2680
  Woodlawn Drive, Honolulu, HI 96822}
\altaffiltext{4}{Harvard Smithsonian Center for Astrophysics,
  60 Garden St., Cambridge, MA 02138}
\altaffiltext{5}{Hubble Fellow}
\altaffiltext{6}{Current address: L--3
        Communications/Brashear, 615 Epsilon Drive, Pittsburgh, PA 15238}
\altaffiltext{7}{US Naval Observatory, 3450 Massachusetts Ave., N.W.,
       Washington, DC 20392-5420}
\altaffiltext{8}{Dept. of Physics and Astronomy, Macquarie University,
  New South Wales, NSW2109, Australia}
\altaffiltext{9}{Australian Astronomical Observatory, P.O. Box 296,
  Epping, NSW 1710, Australia}
\altaffiltext{10}{Gemini Observatory, 670 N. Aohoku Place, Hilo, HI 96720}
\altaffiltext{11}{Current address: Lowell Observatory, 1400 W. Mars
  Hill Rd., Flagstaff, AZ 86001}
\altaffiltext{12}{Alfred P. Sloan Research Fellow}
\altaffiltext{13}{Deceased} 

\begin{abstract}
The white dwarf LHS 3236 (WD1639+153) is shown to be a
double-degenerate binary, with each component having a high mass.
Astrometry at the U.S.  Naval Observatory gives a parallax and
distance of 30.86 $\pm$ 0.25 pc and a tangential velocity of
98 km s$^{-1}$, and reveals binary orbital motion.
The orbital parameters are determined from astrometry of the
photocenter over more than three orbits of the 4.0-year period.
High-resolution imaging at the Keck Observatory resolves the pair
with a separation of 31 and 124 mas at two epochs.  Optical
and near-IR photometry give a set of possible binary components.  
Consistency of all data indicates that the binary is a pair
of DA stars with temperatures near 8000 and 7400 K and with masses
of 0.93 and 0.91 $M_{\sun}$; also possible,
is a DA primary and a helium DC secondary with temperatures near
8800 and 6000 K and with masses of 0.98 and 0.69 $M_{\sun}$.
In either case, the cooling ages of the stars are $\sim$3 Gyr
and the total ages are $<$4 Gyr.  The combined mass of the binary
(1.66--1.84 $M_{\sun}$) is well above the Chandrasekhar limit;
however, the timescale for coalescence is long.

\end{abstract}


\section{Introduction}

A substantial fraction of white dwarfs have binary companions,
often detected as resolved stars with common proper motion.
Deep imaging can find faint companions of white dwarfs,
and high-resolution images can probe toward small separations;
HST has been used this way at visible wavelengths \citep{far10}.
Companions have also been detected as unresolved stars with
redder colors that add to the composite spectral energy
distribution at red and infrared wavelengths.
Infrared imaging from the UKIDSS survey is identifying
the excess flux from unresolved low-mass companions \citep{ste11};
similarly, spectroscopy from SDSS reveals excess red flux from
low-mass companions \citep{sil06}.
Some companions have been detected through radial velocity
variations, despite the difficulty posed by pressure-broadening
of spectral features in white dwarf spectra.  Many companions
found by velocity variability are white dwarfs, but some are
low-mass main sequence companions from the SPY project \citep{nap01}
and from the ELM survey \citep{bro12}.
Some double-white dwarfs are detected 
by the spectral energy distribution being too bright, thus
causing the distance derived from a spectroscopic analysis
to be inconsistent with the measured trigonometric parallax.
In this paper, high-precision astrometry reveals orbital motion
of a previously unknown companion around the white dwarf LHS 3236,
and adaptive optics imaging shows that the binary companion is
also a white dwarf.  In a second paper, similar data will be used
to detect a companion to another white dwarf, G122-31.

Astrometric data are sensitive to detecting companions of low mass
such as brown dwarfs or even massive planets under favorable
circumstances, a potential advantage over radial-velocity detection
of planets around white dwarfs because of the difficulty of measuring
precise radial velocities of white dwarfs mentioned above.
When a perturbation is detected, all of the
orbital elements can be determined, including the inclination,
and there is no uncertainty of ${\rm sin} i$ in masses determined
from the astrometric orbit as there is from a radial velocity
orbit.  One potential complication for astrometric data is that
the orbit of the photocenter is measured, so if the companion
contributes significant light to the photocenter,
then that contribution must be determined and accounted for
in order to determine the true orbits and masses of the components.

The observations used in this paper were taken as part of a
program at the U.S. Naval Observatory to measure trigonometric
parallaxes, distances, and luminosities  of nearby stars.
Of the approximately 200 white dwarfs with accurate parallaxes observed
with CCD cameras at the Naval Observatory over the last 20 years
(Dahn et al. 2013, in preparation), only LHS 3236 and G122-31 have shown
believable perturbations, and the companions in both systems are
found to be white dwarfs.  Nearly a dozen red dwarfs and subdwarfs
observed in this program have perturbations, some with considerably
smaller orbital amplitudes than these two white dwarfs.  Therefore,
the lack of close brown dwarf companions of white dwarfs observed
in this program may be a real effect.

\section{Observations}

LHS~3236 (WD1639+153, PG1639+153, EG 196, G138-56) is a well-known
DA white dwarf, but as yet without a useful parallax measurement.
Its parallax and luminosity would be particularly useful to know
because this star enters the sample of white dwarfs with proper motions
larger than 0.6$''$ yr$^{-1}$ (Luyten 1976; Harris et al. 2001),
important for determining the luminosity function of nearby white dwarfs.
A spectroscopic analysis \citep{ber97}\ gave $T$ = 7450 $\pm$ 360 K
and $M_V$ = 13.29 assuming log $g$ = 8.00, and implying a distance of
30 pc.  Two more recent spectroscopic analyses of DA white dwarfs
include LHS~3236:
a study of white dwarfs in the Palomar-Green survey 
\citep{lie05} yielded $T$ = 7480 K, log $g$ = 8.42 $\pm$ 0.06,
$M_V$ = 13.91, and a high mass of 0.87 $\pm$ 0.04 $M_{\sun}$,
implying a distance of 22 pc;
a survey of bright DA white dwarfs
\citep{gia11} found $T$ = 7550 K, log $g$ = 8.52 $\pm$ 0.07,
$M_V$ = 14.10, and a mass of 0.93 $\pm$ 0.04 $M_{\sun}$,
implying a distance of 21 pc.
The latter survey identified several objects with composite spectra,
but did not detect LHS~3236 as being composite or otherwise unusual.
A spectrum with high resolution \citep{zuc03} found $T$ = 7240 K,
assuming log $g$ = 8.00, with no calcium or other metal lines detected.
A companion tentatively identified \citep{wac03} using the 2MASS Second
Incremental Data Release was based on a $K$ magnitude that is too
bright and is not substantiated in the final 2MASS Point Source
Catalog or in the independent photometry presented below.

New observations presented in the present paper include astrometry
(including its parallax, proper motion, and the discovery of 
perturbation due to binary orbital motion), $BVIJHK$ photometry,
and a measurement of the separation and relative magnitudes of the
two binary components.  The combination of these data allow a nearly
complete determination of properties of the components, with the
result that {\it both} components are massive white dwarfs.
The following subsections describe the observations.

\subsection{Astrometric and Photometric Results}

The astrometry was carried out with the 1.55~m Strand Astrometric Reflector
at the Flagstaff Station of the U.S. Naval Observatory using the
Tektronix 2048x2048 CCD Camera.  The observational and reduction
procedures are those described by Dahn et al. (2002; 2008).
The wide--R filter was used.  Corrections to the astrometry for
differential color refraction for observations taken slightly away
from the meridian were derived from the $V-I$ colors of the parallax
star and the reference stars, as described in Monet et al. (1992).
The correction from relative to absolute parallax was derived
from photometric parallaxes of the individual references stars,
using M$_{\rm V}$ versus $V-I$ relations calibrated with stars with
large trigonometric parallaxes;  this procedure is described fully
in a separate paper (Harris et al. 2013, in preparation).

Photometry with $BVI$ Johnson--Cousins filters was obtained using
the USNO 1~m telescope on three nights, and transformed to standard
magnitudes using Landolt (1992) standard fields to determine nightly
extinction and color terms.  Photometry in $BVI$ taken on one night
is also available \citep{ber97} and is in excellent agreement with
our results.  Photometry in $JHK$ bands was
obtained on one night with the United Kingdom Infrared Telescope
on Mauna Kea, and transformed to the CIT standard system.
Magnitudes in $JHK$ are also available from the 2MASS Point
Source Catalog; however, the faintness of LHS~3236 at infrared
wavelengths makes the 2MASS errors larger than are desirable
(0.04 at $J$ to 0.13 at $K$), so the 2MASS data are not included here.
The combination of $BVI$ and $JHK$ colors gives no indication of a
composite system of two stars with different temperatures.  In fact
the observed K magnitude is slightly fainter than the spectral energy
distribution predicts for a normal DA white dwarf based on the other
photometry.  However, two stars with slightly different temperatures
are not ruled out, as is discussed below.

Astrometric observations of LHS~3236 began in early 1998, and in 2001
residuals from the solution for parallax and proper motion began showing
a significant deviation indicating that this object is an unresolved binary.
To date, 376 acceptable observations have been obtained spanning an epoch
range of 14.3 years and more than three full periods.  These data permit
solutions for the photocentric orbital elements
as well as the parallax and proper motion for this binary system.
As described for another binary, LSR1610$-$00 (Dahn et al. 2008),
an iterative analysis has been adopted, first solving for parallax
and proper motion, then taking the residuals and solving for the orbital
motion of the center of light about the center of mass.  In this paper
(unlike for LSR1610$-$00), we have used the data from the resolved imaging
described below (the position angle and separation) as additional
constraints in the fit for orbital motion.  These preliminary
orbital elements were then used to correct the original astrometry and
repeat the solution for parallax and proper motion and then for orbital
motion.  This iterative process was repeated to convergence.
Table 1 gives the basic astrometric and photometric results.
Plots of the orbital motion (the residuals from the final parallax
solution) are shown in Figure 1 and discussed in Section 4.

Without further information, the orbital solutions from our
astrometry alone are limited in accuracy because the orbital
motion is not far from the noise -- the observed full amplitude
of orbital motion is 7 mas on the sky, while the noise of a single
observation is 3 mas.  Furthermore, the true location of the center
of mass is unknown, requiring a solution for a total of 12 free
parameters.\footnote{
Five free parameters come from the position of the unknown center
of mass (RA and Dec), the proper motion (RA and Dec), and the parallax.
Seven more free parameters come from the orbital elements of the binary
(P, a, e, i, $\Omega$, $\omega$, and T$_{\rm o}$).
}
If based solely on the unresolved astrometric data in Fig. 1,
the possible orbits include a wide range of orbital parameters and
characteristics of the companion.  For example, a massive white dwarf
and a faint substellar companion would be allowed.  Without additional
data, the fractional contribution of the companion to the combined
light is very uncertain, and therefore the size of the true orbit is
poorly known.  With the intention to further constrain the masses
of the white dwarf and its companion and the orbital parameters,
we acquired the high-resolution imaging described in the next subsection.

\subsection{High Resolution Imaging}

\newcommand{\degs}{\mbox{$^{\circ}$}}

We observed LHS~3236 on UT 2008 April 28 and again on UT 2010 May 1
using the laser guide star adaptive optics (LGS AO) system at the
Keck~II Telescope on Mauna Kea, Hawaii \citep{wiz06, vand06}.  The LGS
provided the wavefront reference source for AO correction, with the
exception of tip-tilt motion. The LGS brightness, as measured by the
flux incident on the AO wavefront sensor, was equivalent to a
$V \approx 9.8$--9.9\,mag star. The tip-tilt correction and quasi-static
changes in the image of the LGS as seen by the wavefront sensor were
measured contemporaneously by a second, lower bandwidth wavefront
sensor monitoring LHS~3236, which saw the equivalent of a
$R \approx 15.6$--15.9\,mag star.

On 2008 April 28, data were obtained with the 9-hole non-redundant
aperture mask installed in the filter wheel of the the facility
near-infrared camera NIRC2 \citep{tut06}.  The data were taken in
two dither positions separated by 3$\farcs$5, with
four to six 50~s exposures taken at each dither position. We used the
narrow field-of-view camera and obtained data in $H$- and $K_S$-bands.
Typical interferograms from our observations are shown in
Figure~\ref{fig:masking}.  The pipeline used to reduce the aperture
masking data was similar to that used in previous papers containing
NIRC2 masking data \citep{ire08, kra08, irek08},
except that no comparable single star was observed in $K_S$-band for
calibration of the closure phases or squared visibilities. Thus we
chose to fit only the $K_S$-band closure phases with a binary
model, whereas for the $H$-band data we fit the calibrated closure
phases and squared visibilities. For the model fit, the closure phase
uncertainties were initially approximated by the standard error of the
mean calculated from the scatter among individual exposures. The
uncertainties were subsequently increased by adding a calibration
error in quadrature so that the resulting reduced $\chi^2$ of the fit
was 1.0. Although we fit all 84 closure phases from the 9-hole mask,
only 28 of these are formally independent. To correctly account for
this non-diagonal covariance matrix in our binary fitting, we scaled
the errors in the least-squares fit to the data by
$\sqrt{84/28}$. This process has been validated both by a comparison
to fits using full covariance matrices \citep{kra08} and
by orbit fits using mixed data that resulted in a reduced $\chi^2$
consistent with unity, where the orbit fit had many degrees of freedom
\citep[e.g.,][]{mar07}.

The closure phases resulting from our binary model fits to the data
are shown in Figure~\ref{fig:masking}. Because LHS~3236 is a nearly equal
magnitude binary, it predominantly has closure phases close to 0\degs\
or $\pm$180\degs. In 2008, the binary is clearly detected in both filters,
though the $H$-band data have more closure phases in which the binary
is detected, and thus the binary parameters are much better determined
from the $H$-band data. They give a separation of 30.5$\pm$0.2~mas,
a position angle of 276.1$\pm$0.8\degs, and a flux ratio of
0.090$\pm$0.016~mag.  This implies that the binary separation is near
the resolution limit in $K_S$ band, where almost all the information
is carried by a single baseline, and thus the formal errors may be
underestimated.  In spite of this, we find a binary solution in $K_S$
band with reasonably consistent separation and P.A. ($31.8\pm0.3$\,mas
and $278\fdg1\pm0\fdg6$) and a flux ratio of $0.10\pm0.04$\,mag).

On 2010 May 1, we obtained direct imaging at $K_S$ band without the
aperture mask.  At this epoch the binary separation was actually
expected to be much wider and thus near the outer limits of the
masking field of view of $\approx$125\,/mas.  To reduce and analyze
our images we used the same methods as in our previous work
\citep{dup09a, dup09b, dup10}, fitting three-component
elliptical Gaussians to measure binary parameters.  Figure~3 shows
one of our four $K_S$-band images, from which we determined a binary
separation of $123.9\pm0.4$\,mas, a P.A.\ of $88\fdg32\pm0\fdg13$, and
a flux ratio of $0.005\pm0.013$\,mag.  The flux ratio agrees with our
earlier masking data, and in the following analysis we adopt the
$K_S$-band flux ratio from imaging since its uncertainty is expected
to be more reliable than for the earlier $K_S$-band masking data.
Note that in all of the above analysis we have used the
\citet{ghe08} calibration of NIRC2, with a pixel scale
of $9.963\pm0.005$\,mas/pixel and orientation of the $+y$-axis of
$+0\fdg13$.

\section{Photometric and Spectroscopic Analysis}

The parallax in Table 1 gives a distance of 30.86 $\pm$ 0.25 pc,
larger than the spectroscopic distance of 21-22 pc \citep{lie05, gia11},
and therefore could allow a nearly normal mass for LHS 3236 rather than
the high mass found in that study.
Surprisingly, however, the nearly equal magnitudes of the two components
requires that LHS 3236 be a double-degenerate binary, eliminating the
possibility of a substellar companion.  Furthermore, because each component
is fainter than the combined light by 0.7-0.8 mag at $H$ and $K$,
then {\it both} components must be more massive than normal white dwarfs.

The photometry of LHS~3236 plus the parallax given in Table 1
give the absolute magnitudes in each filter of the combined light
of the two components of the binary.  With the additional information
from Sec. 2.2 that the two components differ by 0.09 mag at $H$ and
0.01 mag at $K$, we can draw conclusions about the two components
from the photometry and parallax alone, under certain assumptions.

The simplest possibility is that the two components are DA white
dwarfs with the same temperature.  Under this assumption, a fit
of the photometry to the models of DA white dwarfs from
\cite{ber95}\footnote{
The model colors are taken from the tables at
www.astro.umontreal.ca/$\sim$bergeron/CoolingModels.}
gives a temperature of 7700 $\pm$ 60 K.  The deconvolved component
absolute magnitudes are given in the first two lines of Table 2.
The surface gravities and masses of each component also can be
determined from the Bergeron models, and are given in Table 2.
Finally, using these masses and magnitudes, the dilution of the
semi-major axis of the true relative orbit $a$ to the smaller
observed semi-major axis of the photocenter orbit $\alpha$ can be
determined from
\begin{displaymath}
  \alpha = a (f - \beta)
\end{displaymath}
where
\begin{displaymath}
  f = M_2/(M_1 + M_2)
\end{displaymath}
is the fractional mass of the secondary, and
\begin{displaymath}
\beta = l_2/(l_1+l_2) > 0
\end{displaymath}
is the fractional light of the secondary in the $R$-band
astrometric filter.  These values are also given in Table 2.

The photometry does not require that the two stars have the same
temperature, and two additional possible combinations of DA+DA
components are given in Table 2, with an increasing temperature
difference between the two stars.  The combined light of the pairs
of model DA stars given in Table 2 agrees with observed
combined-light magnitudes from Table 1 within 0.03 mag in each
filter.  These three pairs in Table 2 are examples drawn from an
infinite set of possible pairs with a continuous distribution of
possible temperature differences:  for each chosen example,
the absolute magnitudes and masses of both stars given in the
table are the only possibilities (within a small range allowed
by observational errors discussed below) that reproduce the observed
combined-light magnitudes and the observed H-band magnitude
difference.  However, in Section 4, we find that the first
pair with equal-temperature stars, and other pairs with larger
temperature differences, are not possible for dynamical reasons.

It can be seen in Table 2 that all the DA+DA binaries have log~$g$
close to 8.5 and a mass close to 0.9 M$_{\sun}$ for each component.
This gravity is in excellent agreement with the spectroscopic
determinations of 8.42 and 8.52 \citep{lie05, gia11} of the combined light.
The temperatures near 7700 K are slightly warmer than the spectroscopic
values 7480 and 7550 K, primarily because the {JHK} photometry
(which was not available to Liebert et al. or Gianninas et al.)
shows that the pair is fainter than would be the case at the cooler
temperature.

The photometry also allows the fainter component of the binary to be
a DB or DC (helium atmosphere) white dwarf.  For the cool temperature
that any helium atmosphere companion must have, no helium lines would
be visible, so we will refer to these as DA+DC pairs.  The second half
of Table 2 gives three examples of possible DA+DC pairs with an increasing
difference in temperature between the two stars.
The DA+DC pairs in Table 2 agree with the photometry in Table 1
within 0.02 mag in each filter.  For all of these DA+DC pairs,
the Balmer absorption lines from the DA star would be made weaker
in the combined-light spectrum by the added continuum of the DC star.
For these combinations, the DA can have a warmer temperature than for
the DA+DA pairs.  At temperatures near 9000 K, the strong Balmer
absorption lines from the DA will be diluted by the DC.

The spectrum of LHS~3236 from \citet{gia11} (kindly made available
to us by Bergeron), is shown in Figure 4, compared with two composite
spectra of our DA+DA and DA+DC pairs, based on model spectra from
\citet{koe10} (kindly made available to us by Koester).
Both pairs are consistent with the observed spectrum, although the
DA+DC pair tends to have all the Balmer lines fit better by the models,
whereas the models of the DA+DA pair show H$\beta$ stronger than
observed and H$8$ weaker.  (The DA+DC pairs from Table 2 with a
hotter DA component at 8800 or 9000 K do not fit the observed spectrum
as well, because the high Balmer lines become too strong in those models.)
The spectrum observed by \citet{zuc03}\footnote{
This spectrum was taken on 1999 Apr 19 UT.  The orbital period of
4.03 yr implies the binary was then at the same orbital phase as in
July 2011, and the predicted radial velocity curves would have a
velocity difference of only 8 km s$^{-1}$.  Therefore, the spectrum
could not have shown resolved Balmer lines from the two stars.
}, with higher spectral resolution,
suggests the same conclusion, because the high Balmer lines
appear strong in the observed spectrum.  However, the lack of
flux calibration for the archived spectrum affects the high Balmer
lines H$\epsilon$ and H$8$ and makes this conclusion less definitive.

The conclusion of this section is that pairs of DA+DA white dwarfs
with masses near 0.9 M$_{\sun}$ each are consistent with the available
photometry, spectra, and parallax, and the combined mass is 1.84 $\pm$ 0.03
M$_{\sun}$ for two DA components.  Alternatively, pairs of DA+DC white
dwarfs are also possible with masses near 1.0 M$_{\sun}$ for the
DA component and 0.7 M$_{\sun}$ for the DC component, and with a
combined mass of 1.68 $\pm$ 0.05 M$_{\sun}$.
The next section discusses the observed orbit to look for consistency
with this photometric analysis.

\section{Dynamical Analysis}

The orbital motion of the photocenter shown in Fig. 1,
with the separation and position angle from the high-resolution
imaging in Section 2.2, is best fit by the curve shown in Fig. 1.
The orbit projected onto the plane of the sky is shown in Fig. 5,
with the two observed positions of the secondary star marked.
The orbital elements of the photocenter are given in Table 3.

The masses from Table 2 and the period from Table 3 can be used
with Kepler's third law
\begin{displaymath}
 { a{^3}/P^2 = M_1+M_2 }
\end{displaymath}
to calculate the semi-major axis $a$ = $a_1 + a_2$ of the relative
orbit.  The result is 3.10 or 3.00 AU using the masses for DA+DA
or DA+DC pairs, respectively.
The semi-major axis of the photocenter orbit $\alpha$
in Table 3 is 4.37 mas, or 0.135 AU.  These require the dilution factor
of the photocenter amplitude $a/\alpha$ to be 23.0 or 22.2, respectively.
This large dilution factor is consistent with the observed motion
of the resolved secondary star between 2008 and 2010 (31 mas west
in 2008 to 124 mas east in 2010) compared to the observed motion
of the photocenter (roughly 1.4 mas east to 5.4 mas west).
The temperatures of the pairs of white dwarfs given in Table 2
are chosen to satisfy this dilution factor (except the first pair
of equal-temperature DA stars, shown in Table 2 as an example
of a pair not consistent with this constraint).
A DA+DA binary, with a combined mass of 1.84 M$_{\sun}$ from Table 2,
agrees with the orbital parameters in Table 3, the dilution factor
calculated in Table 2, and the observed separations plotted in Fig. 5.
A DA+DC binary, with a combined mass of 1.68 M$_{\sun}$ from Table 2,
is possible, but must have a photocenter orbit with smaller $\alpha$.
The fits to the data in Figures 1 and 5 tend to give slightly better
fits with larger $\alpha$ (usually accompanied by larger eccentricity),
and slightly poorer fits with $\alpha$ as small as 3.00 AU.
Therefore, these data indicate that a DA+DA binary is most likely,
but a DA+DC binary is not ruled out.  The spectroscopic data (Sec. 3)
suggested a DA+DC binary is more likely, and we conclude that
either possibility is acceptable.

The error in the combined mass has contributions from the errors
in the parallax and photometry and the uncertainty in the temperatures
of the two stars, but the largest uncertainty is DA or DC type of
the secondary component.  In either case,
the combined mass is well above the Chandrasekhar limit.
With a relative semi-major axis of 3.1 AU, the separation between
the components at periastron is 0.8 AU.  This separation is large
enough that the timescale to reduce the size of the orbit through
gravitational radiation and lead toward coalescence and a supernova
explosion is longer than the age of the universe.

Some insight into the evolution of the system can be obtained from the
white dwarf cooling age of each star.  The models of DA and DC white
dwarfs \citep{ber95} used for Table 2 give a cooling age near 3.0 Gyr
for each component, and nearly the same for a DA or DC secondary.
Using the initial-final mass relation from \cite{wil09}, the progenitor
stars would each have had masses of approximately 4.5 M$_{\sun}$.
Stellar evolution models (e.g. Bertelli et al. 2008) give progenitor
lifetimes for main-sequence to AGB evolution of 0.3 Gyr.
Therefore, the total age of the system is $\sim$3.3 Gyr.

It is likely that the B component evolved first.
Significant mass may have been transferred to the progenitor of A
during the AGB evolution of B, or the mass lost by B may have
been lost from the system.  This unknown factor makes the initial
orbit and the initial mass of A also unknown.  Mass lost from A
during its AGB evolution also could have been transferred back to B
or lost from the system, and would have further modified the orbit.
The evolutionary models \citep{ber08} give a maximum radius for
a 4.5 M$_{\sun}$ progenitor of 3 AU during AGB evolution,
equal to the present semi-major axis, and probably not a coincidence.
Therefore, mass loss by Roche lobe overflow is expected.
Mass loss correlated with the orbital phase, and possibly also
with pulsation of the AGB star, could act to increase the eccentricity
to the large value we see now.  It is likely that an interesting
asymmetric planetary nebula would have been created {\it twice}
around this system as each star passed through its post-AGB stage.

\section{Future Observations}

Several observations are possible to confirm the results of this
paper and to further constrain the components of this binary system.
First, astrometry during the next periastron passage late in the 2016
observing season (both optical astrometry of the photocenter and
high-resolution astrometry of the resolved pair) can further define
the rapidly changing orbital curve.
Second, a high-resolution measure of the magnitude difference
between the components at an optical band like $V$ can be done with
HST to confirm the temperature difference between the two components.
Third, a determination of the radial velocity curve of the primary
star can be done with high signal-to-noise spectroscopy such as the
SPY program achieves to help constrain the components.
Figure 6 shows the predicted velocity curve for the most likely orbit.
The rapid change in velocity of 20--30 km s$^{-1}$ near periastron
may be measureable.  Also, it may be feasible to resolve
absorption lines of the secondary star if it is a DA white dwarf,
a measurement that would provide a determination of the masses of
the components independent of the interior models used in this paper.
Fourth, a high signal-to-noise optical spectrum would give an
improved spectroscopic estimate of the mass(es), particularly
when combined with the parallax and photometry in this paper.

\acknowledgments

We thank the referee for a constructive review, and suggesting
comparison of our binary pairs with archival spectra.
We thank A. Gianninas and P. Bergeron for making their unpublished
spectrum available to us, and we thank D. Koester for making his
model spectra available.
This research has made use of the USNOFS Image and Catalogue Archive
operated by the United States Naval Observatory, Flagstaff Station
(http://www.nofs.navy.mil/data/fchpix).
We acknowledge the Keck AO team for their exceptional efforts in
bringing the AO system to fruition. We thank Al Conrad, Jason McIlroy,
Gary Punawai, Hien Tran, and the Keck Observatory staff for assistance
with the observations.  We acknowledge the significant contribution of
Peter Tuthill in his work to establish aperture masking at Keck.
T.J.D.\ acknowledges support from Hubble Fellowship grant
HST-HF-51271.01-A awarded by the Space Telescope Science Institute,
which is operated by AURA for NASA, under contract NAS 5-26555.
T.J.D.\ and M.C.L.\ acknowledge support for this work from NSF grants
AST-0507833 and AST-0909222. M.C.L.\ acknowledges support from an
Alfred P.\ Sloan Research Fellowship.
Finally, the authors wish to recognize and acknowledge the very
significant cultural role and reverence that the summit of Mauna Kea has
always had within the indigenous Hawaiian community.  We are most
fortunate to have the opportunity to conduct observations from this
mountain.

\clearpage

\begin{deluxetable}{lr}
% \tabletypesize{\scriptsize}
% \tabletypesize{\footnotesize}
\tablecolumns{2}
\tablewidth{0pt}
\tablecaption{Astrometric and Photometric Results for LHS 3236(A+B)}
\tablehead{
\colhead{Result} &
\colhead{Value} }
\startdata
Epoch Range (yr)                        &             14.3  \\
No. Frames                              &              376  \\
No. Nights                              &              287  \\
No. Ref. Stars                          &               14  \\
Rel. Parallax (mas)$^1$                 & 32.40 $\pm$ 0.33  \\
Rel. Parallax (mas)$^2$                 & 31.67 $\pm$ 0.25  \\
Rel. Proper Motion (mas yr$^{-1}$)      & 668.61 $\pm$ 0.04 \\
P.A. of Proper Motion (degrees)         & 178.83 $\pm$ 0.01 \\
Correction to Abs. Parallax (mas)       &  0.74 $\pm$ 0.06  \\
Abs. Parallax (mas)$^2$                 & 32.41 $\pm$ 0.26  \\
V$_{\rm{tan}}$ (km s$^{-1}$)            &  97.8 $\pm$ 0.8   \\
V                                       & 15.70 $\pm$ 0.02  \\
B$-$V                                   &  0.32 $\pm$ 0.02  \\
V$-$I                                   &  0.45 $\pm$ 0.02  \\
J                                       & 15.08 $\pm$ 0.04  \\
H                                       & 14.95 $\pm$ 0.05  \\
K                                       & 14.94 $\pm$ 0.05  \\
M$_{\rm{V}}$                            & 13.25 $\pm$ 0.03  \\
\enddata
\tablenotetext{1}{Before removal of the astrometric perturbation}
\tablenotetext{2}{After removal of the astrometric perturbation}
\end{deluxetable}
\clearpage

\begin{deluxetable}{llrrrrrrrrrrr}
% \tabletypesize{\scriptsize}
% \tabletypesize{\footnotesize}
\tablecolumns{13}
\tablewidth{0pt}
\tablecaption{Binaries Consistent With Photometry}
\tablehead{
\colhead{Component} &
\colhead{T$_{\rm eff}$} &
\colhead{M$_B$} &
\colhead{M$_V$} &
\colhead{M$_R$} &
\colhead{M$_I$} &
\colhead{M$_J$} &
\colhead{M$_H$} &
\colhead{M$_K$} &
\colhead{log g} &
\colhead{Mass} &
\colhead{{$\beta$}(R)} &
\colhead{a/$\alpha$} }
\startdata
\multicolumn{4}{l}{DA+DA Binaries}&&&&&&&&& \\
A& 7700& 14.28& 13.98& 13.78& 13.54& 13.38& 13.21& 13.21& 8.49& 0.910& 0.487& 54.8 \\
B& 7700& 14.33& 14.04& 13.84& 13.60& 13.43& 13.26& 13.26& 8.52& 0.930&      &      \\
A& 8000& 14.16& 13.89& 13.71& 13.50& 13.36& 13.21& 13.22& 8.53& 0.932& 0.453& 23.8 \\
B& 7400& 14.47& 14.14& 13.92& 13.66& 13.46& 13.27& 13.27& 8.49& 0.912&      &      \\
A& 8100& 14.14& 13.87& 13.70& 13.49& 13.36& 13.22& 13.23& 8.54& 0.942& 0.447& 23.9 \\
B& 7300& 14.50& 14.16& 13.93& 13.67& 13.46& 13.26& 13.25& 8.48& 0.900&      &      \\
\multicolumn{4}{l}{DA+DC Binaries}&&&&&&&&& \\
A& 8600& 13.96& 13.72& 13.58& 13.40& 13.32& 13.21& 13.23& 8.58& 0.970& 0.388& 24.4 \\
B& 6200& 14.91& 14.40& 14.07& 13.74& 13.51& 13.36& 13.27& 8.24& 0.728&      &      \\
A& 8800& 13.90& 13.68& 13.54& 13.38& 13.32& 13.21& 13.24& 8.61& 0.985& 0.372& 23.7 \\
B& 6000& 15.02& 14.47& 14.11& 13.76& 13.50& 13.34& 13.25& 8.19& 0.695&      &      \\
A& 9000& 13.85& 13.63& 13.51& 13.36& 13.32& 13.22& 13.25& 8.63& 1.000& 0.355& 24.3 \\
B& 5800& 15.15& 14.54& 14.16& 13.78& 13.49& 13.31& 13.21& 8.13& 0.655&      &      \\
\enddata
\end{deluxetable}
\clearpage

\begin{deluxetable}{lr}
% \tabletypesize{\scriptsize}
% \tabletypesize{\footnotesize}
\tablecolumns{2}
\tablewidth{0pt}
\tablecaption{Orbital Parameters for LHS 3236}
\tablehead{
\colhead{Parameter} &
\colhead{Value} }
\startdata
Period (yr)    &   4.030 $\pm$ 0.018 \\
$\alpha$ (mas)$^1$& 4.37 $\pm$ 0.25 \\
$\alpha$ (AU)$^1$ &0.135 $\pm$ 0.008\\
i (deg)        &    93.2 $\pm$ 0.3 \\
e              &   0.740 $\pm$ 0.032 \\
$\omega$ (deg) &    49.4 $\pm$ 1.6  \\
$\Omega$ (deg) &    91.6 $\pm$ 0.3  \\
T$_0$          & 2008.54 $\pm$ 0.03 \\
RMS$_{\rm RA}$ (mas)  &  3.0 \\
RMS$_{\rm Dec}$ (mas) &  3.4 \\
\enddata
\tablenotetext{1}{Semi-major axis of the photocentric orbit.}
\end{deluxetable}

\clearpage
\begin{figure}
\plotone{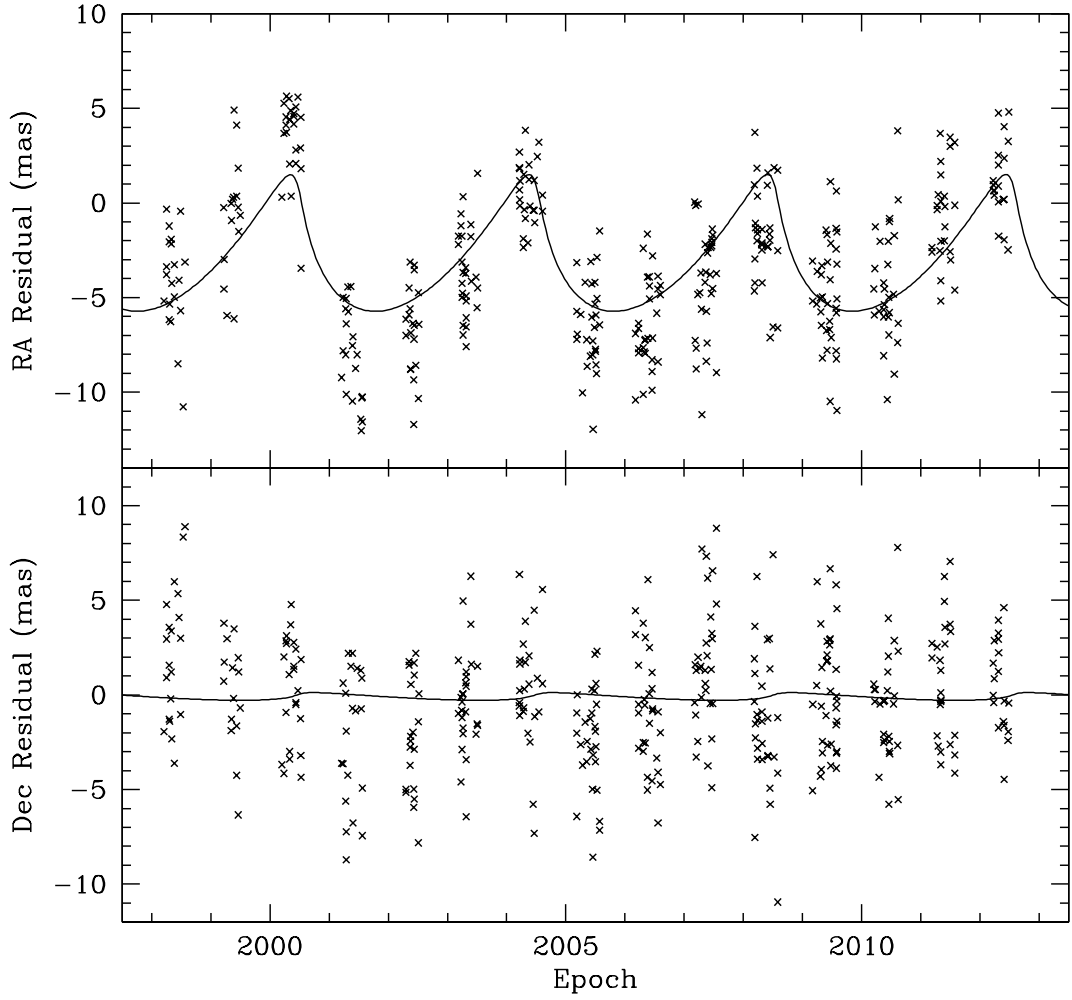}
\caption{The positions of the combined light of LHS~3236
in Right Ascension (upper panel) and Declination (lower panel),
in mas, after removing the parallactic and proper motion.
\label{Fig.1}}
\end{figure}

\clearpage
\begin{figure} 
\centerline{\includegraphics[width=5.0in,angle=0]{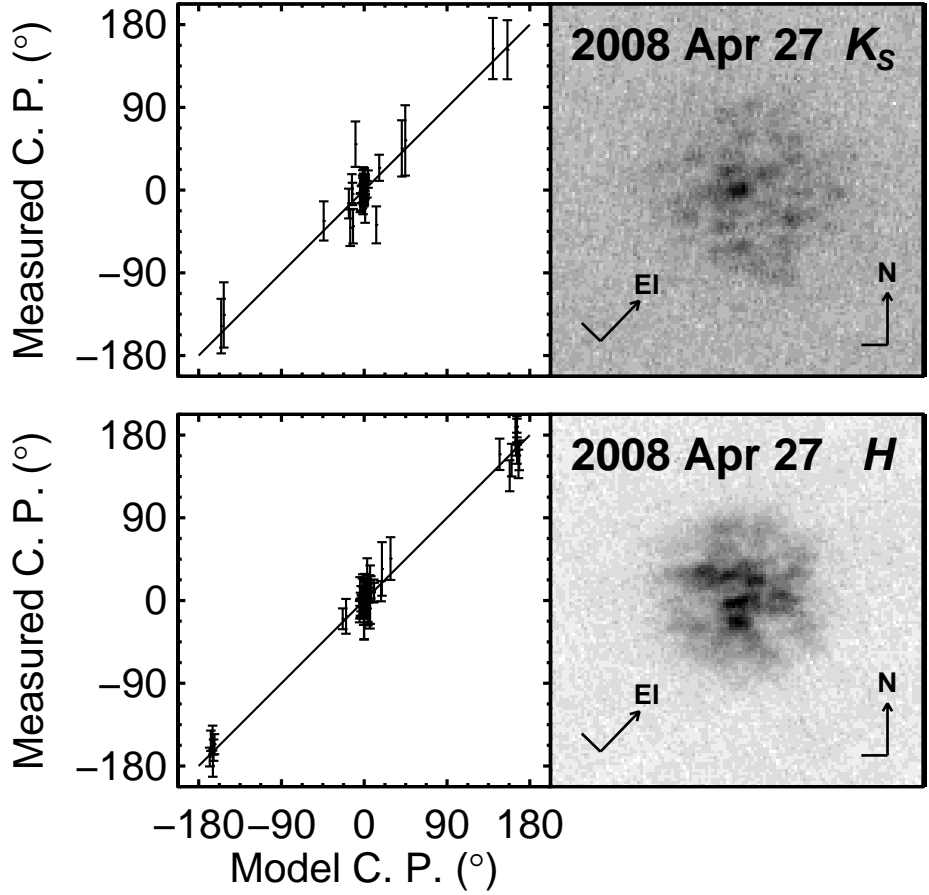}}
\caption{ \normalsize \emph{Right:}
  Keck interferograms of LHS~3236 obtained in 2008 using NIRC2's
  9-hole aperture mask (square-root stretch).  The direction north
  is labeled N, the elevation axis is labeled El.  Individual
  spots in these images correspond to closure triangles of different
  baselines \citep[e.g., see][]{tut06}.  These spots show a slight
  elongation in the direction of P.A. of the binary ($\approx$276\degs).
  (The elongation would be in the elevation direction if it were an
  effect of refraction or windshake.)
  \emph{Left:} Measured versus modeled closure phases (C.P.), from
  which the binary parameters were derived (we also used the squared
  visibilities for the $H$-band data).  The errors in the binary
  parameters were assessed in a Monte Carlo fashion that accounted for
  the measured closure phase and squared visibility errors.  With
  closure phases only at $\approx$0\degs and $\approx$180\degs,
  a binary with a near unity flux ratio is indicated. \label{fig:masking}}
\end{figure}

\clearpage
\begin{figure} 
\centerline{\includegraphics[width=4.5in,angle=0]{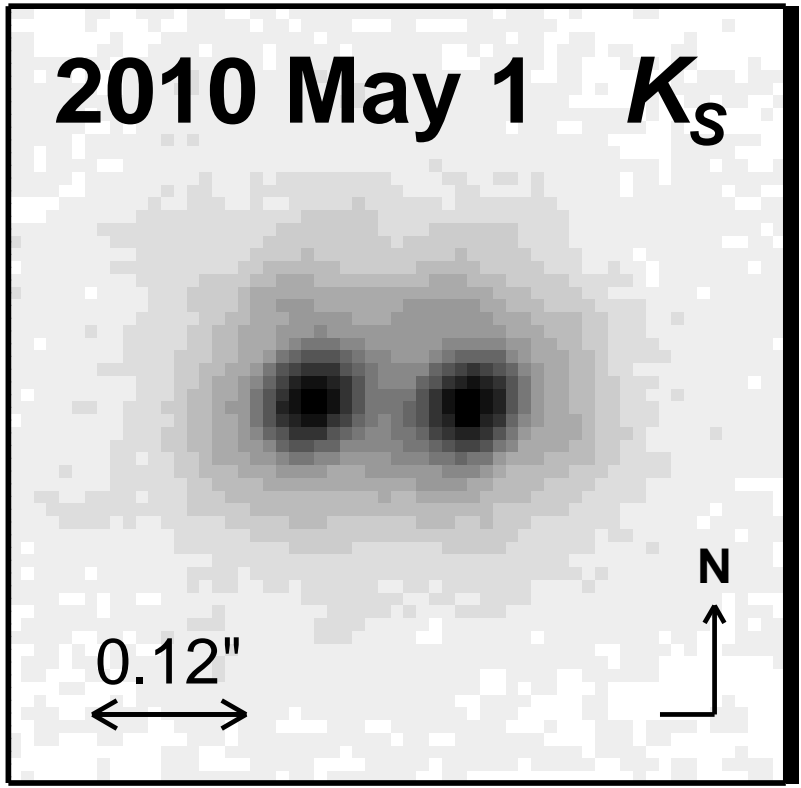}}
\caption{ \normalsize One of four Keck images in $K_S$-band of LHS~3236
obtained in 2010. \label{Fig.3}}
\end{figure}

\clearpage
\begin{figure} 
\plotone{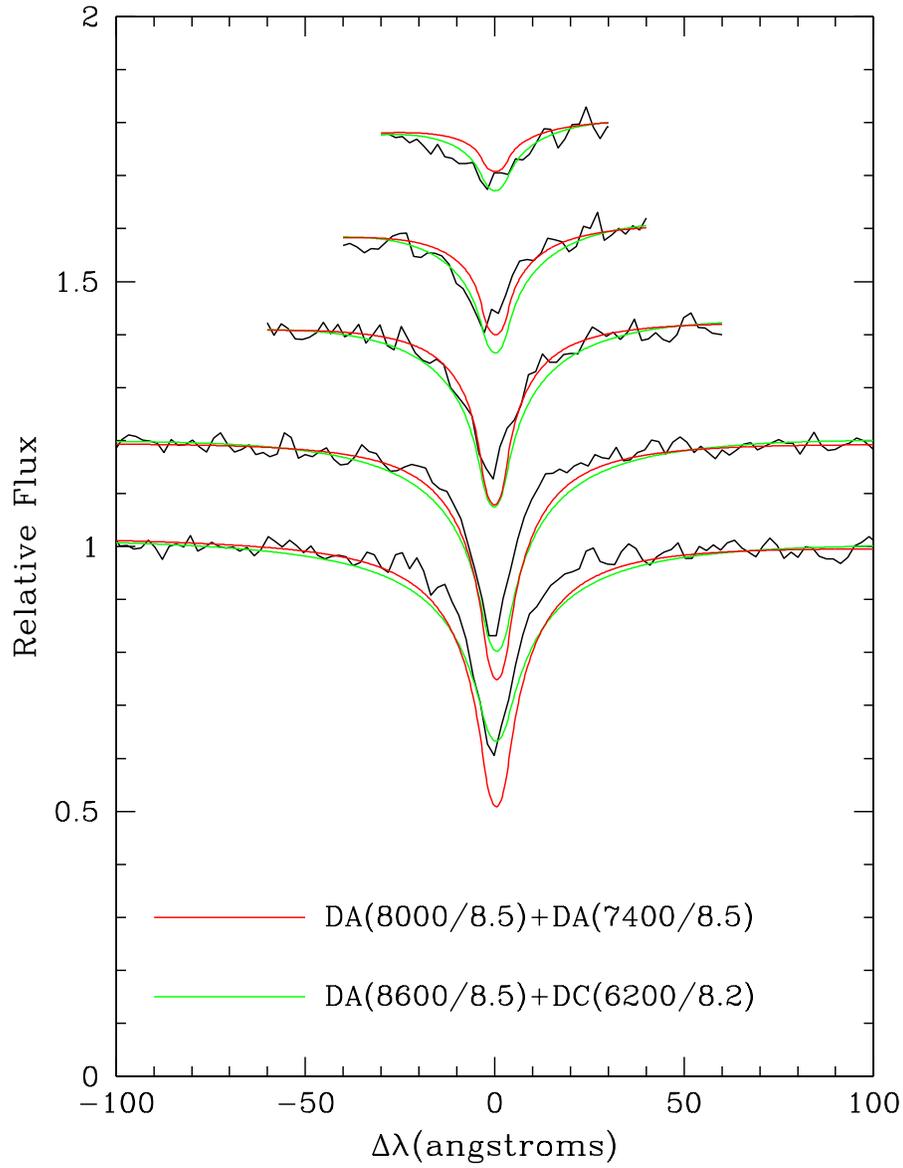}
% \plotone{balmerplotbw.eps}
\caption{The spectrum of LHS~3236 from Gianninas et al. (2011),
compared to two composite spectra from models by Koester (2010).
See the online paper for a color version of this plot.
\label{Fig.4}}
\end{figure}

\clearpage
\begin{figure}
\plotone{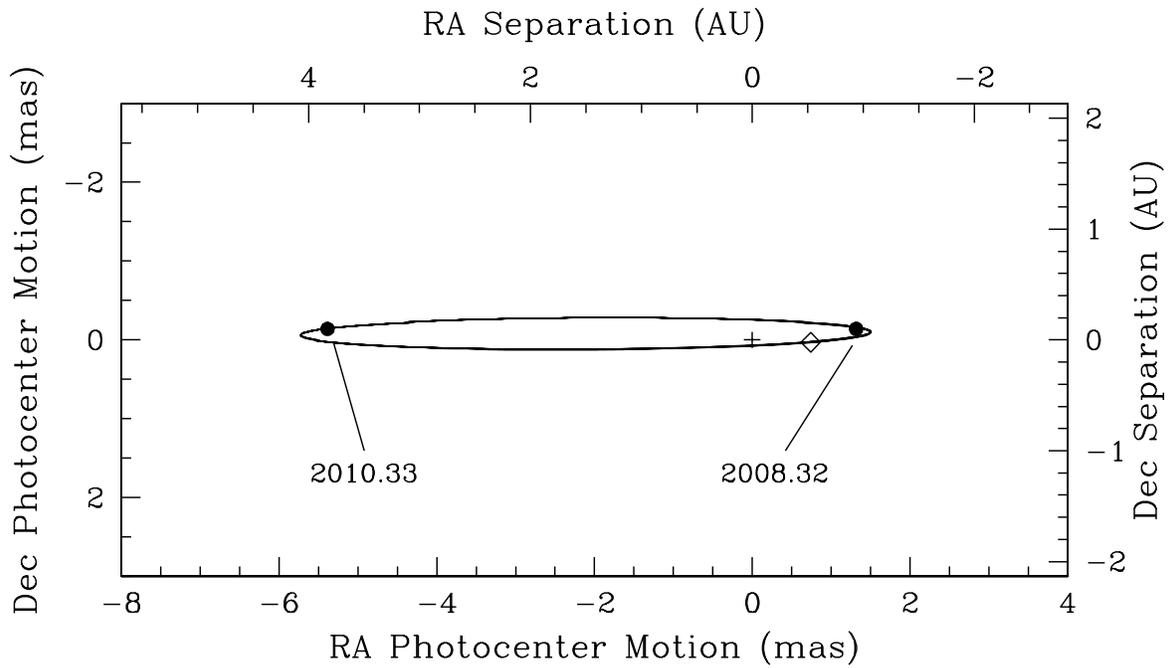}
\caption{The orbit shown on the plane of the sky.
The bottom and left axes indicate the motion of the photocenter
around the center of mass, using the orbital elements in Table 3.
The top and right axes indicate the position of the secondary star
relative to the primary star on the plane of the sky.
The filled circles show the two observations of the companion
seen in the high-resolution images.  The open diamond
next to the origin shows the companion at periastron.
\label{Fig.5}}
\end{figure}

\clearpage
\begin{figure}
\plotone{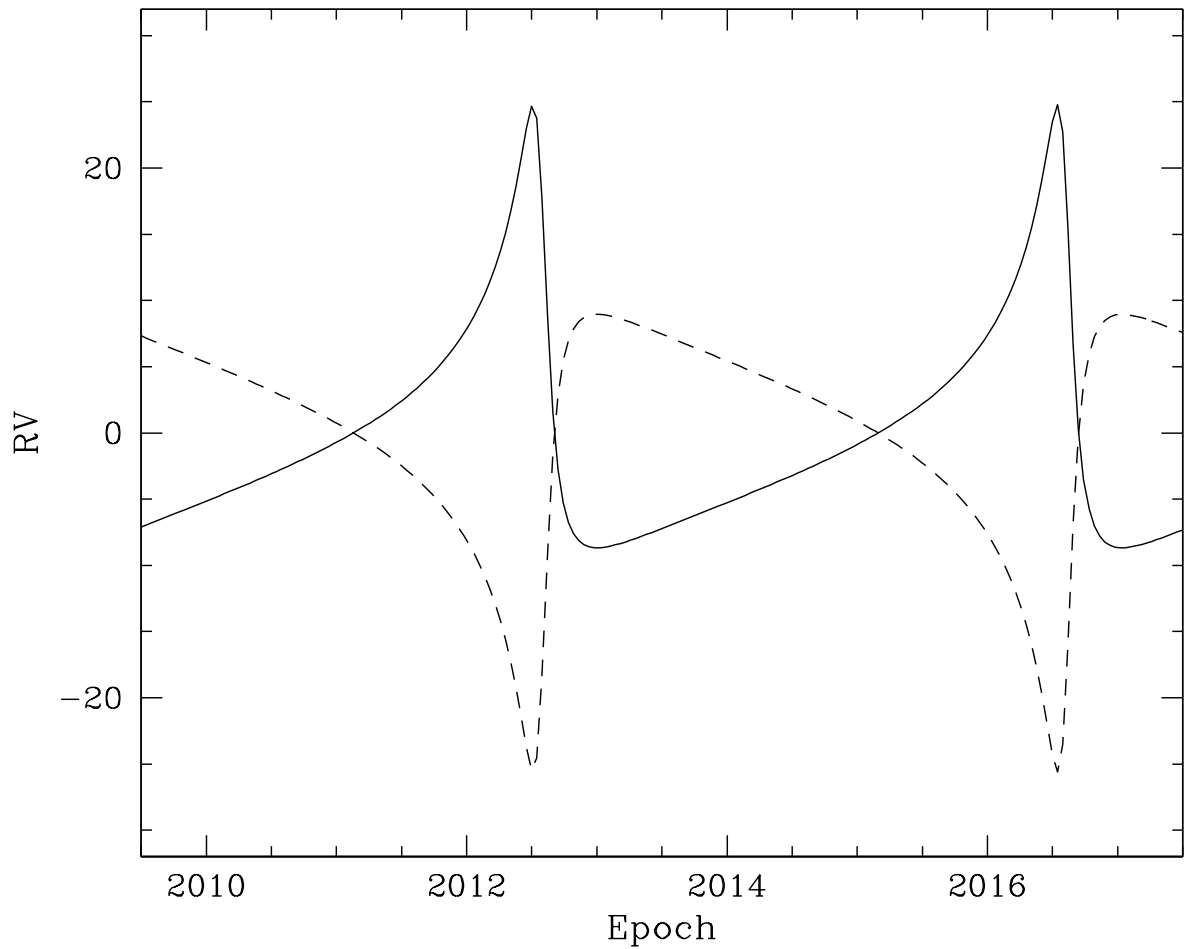}
\caption{The radial velocity curves predicted
for the derived orbit given in Table 3.
The solid curve shows the primary star, and the
dashed curve shows the secondary star for two periods.
Note that the sign of the predicted velocity curves is unknown
and can be changed to match observations.
\label{Fig.6}}
\end{figure}

\end{document}